\documentclass[a4]{article}

\pdfoutput=1 
\usepackage[english]{babel}
\usepackage{amsmath}
\usepackage{amssymb}
\usepackage{theorem}
\usepackage[applemac]{inputenc}
\usepackage{afterpage}
\usepackage[retainorgcmds]{IEEEtrantools}
\usepackage{graphicx}
\usepackage{subfigure}
\graphicspath{{figss/}}
\usepackage{enumerate}
\usepackage{multicol}
\usepackage{url}
\usepackage{float}

\title{The OpenPicoAmp : an open-source planar lipid bilayer amplifier for hands-on learning of neuroscience}
\author{Vadim Shlyonsky$^{1}$, Freddy Dupuis$^{2}$ \& David Gall$^{3}$ \\
\small{$^1$Laboratoire de Physiologie et Physiopathologie, Université Libre de Bruxelles,Bruxelles,  Belgium.}\\
\small{$^2$Service Ondes et Signaux, Université Libre de Bruxelles, Bruxelles, Belgium.}\\
\small{$^3$Laboratoire de Neurophysiologie, ULB Neuroscience Institute, Université Libre de Bruxelles,Bruxelles,  Belgium.}\\}
\date{}

\begin{document}

\maketitle

\begin{abstract}
Neuroscience education can be promoted by the availability of low cost and engaging teaching materials. To address this issue, we developed an open-source lipid bilayer amplifier, the OpenPicoAmp, which is appropriate for use in introductory courses in biophysics or neurosciences at the undergraduate level, dealing with the electrical properties of the cell membrane. The amplifier is designed using the common lithographic printed circuit board fabrication process and off-the-shelf electronic components. In addition, we propose a specific design for experimental chambers allowing the insertion of a commercially available polytetrafluoroethylene film. We provide a complete documentation allowing to build the amplifier and the experimental chamber. The students hand-out giving step-by step instructions to perform a recording is also included. Our experimental setup can be used in basic experiments in which students monitor the bilayer formation by capacitance measurement and record unitary currents produced by ionic channels like gramicidin A dimers. Used in combination with a low-cost data acquisition board this system provides a complete solution for hands-on lessons, therefore improving the effectiveness in teaching basic neurosciences or biophysics.  
\end{abstract}



\section{Introduction}

The traditional lecture is still the standard pedagogical method for teaching science at the undergraduate level, although it has been shown that more active approaches are more efficient especially in large-enrollment courses  \cite{Haak2011, Freeman2014}. In addition, it is now challenged by the development of massive open online course, or MOOC. This evolution should reinforce the interest in extensive hands-on learning sessions as they provide a way to improve learning  which cannot be obtained by the online methods \cite{Deslauriers2011}. Hands-on learning sessions are also relevant in the "flipped classroom" approach which is an inverted teaching structure where instructional content is delivered outside class, and engagement with the content is done in class, under teacher guidance and in collaboration with peers. Instead of giving the same explanations over and over, the teacher can capture his explanation once on video or audio, and spend energy and time individualizing instruction \cite{Prober2012}. This provides a way to cope with large class sizes and reach students who are at varying levels of understanding and skill.

In undergraduate basic science courses, the hands-on learning sessions can take the form of laboratory sessions where of a series of challenging questions and tasks require students, divided in small groups, to practice reasoning and problem solving while provided with frequent feedback from fellow students and from the instructors. To implement this approach there is crucial need of engaging teaching materials. Entry-level neurophysiology equipment used for teaching neuroscience are in the range of thousands of euros and experiments may require the use of living animals. Here we propose an open-source lipid bilayer amplifier which is appropriate for use in introductory courses in biophysics or neuroscience. We also describe a simple experimental protocol which is currently performed by undergraduate medical students during their training in basic neuroscience at our institution. The related costs to build the amplifier and the bilayer chamber are below 200 euros and the experiments do not involve the use of animals.


One way to show that you understand how something works is to build it. As we are able reconstitute a neuronal cell membrane at the molecular level using the so-called black lipid membrane technique (BLM)  \cite{MUELLER1962}, we can say that we understand how neurons produce electrical signals. The planar lipid bilayer method allows to build an artificial cell membrane and examine pore forming molecule functionally at the single-molecule level. The laboratory session we propose in this paper allows the students to observe directly these elementary currents produced by the activity of a single pore forming molecule, such ionic channels being the key players in the generation of the neuronal electrical activity.

\section{Materials and Methods}

\begin{figure}
\centering
\includegraphics[width=0.7\linewidth]{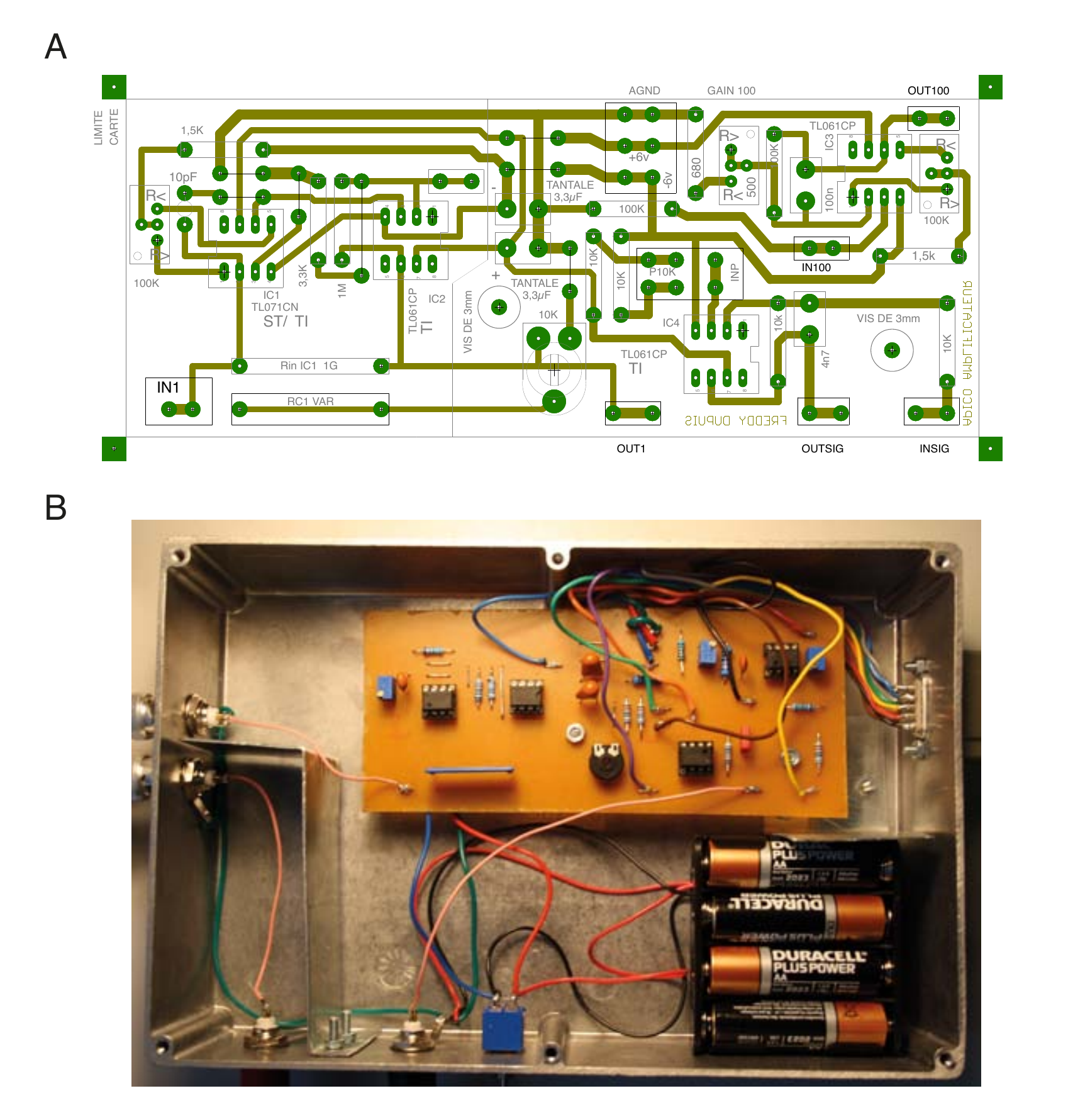}
\caption{The OpenPicoAmp amplifier. The chosen electronic design uses a minimal number of commonly found components, like the Texas Instruments TL061 and TL071 operational amplifiers. It is therefore inexpensive and easy to build. (A) Printed circuit board layout for the OpenPicoAmp. (B) Implantation of the circuit and two 6 $V$ batteries in the main box. An additional box is necessary to build the remote control allowing to switch gain and adjust the offset without producing vibrations that may compromise the stability of the lipid bilayer (not shown).}
\label{OpenPicoAmp-fig}
\end{figure}

\subsection{Participants}
The laboratory session is typically lasting four hours and starts with a brief 15-20 minutes lecture focusing on the electrical properties of the cell membrane and the voltage clamp technique. Students are reminded that an element of passive cell membrane (i.e. not producing action potentials) is electrically equivalent to a $RC$ circuit. In addition, the operation of the voltage clamp circuit, which is an op-amp current to voltage converter circuit, is explained. The latter allows us to introduce the general concept of negative feedback which is relevant in many biological processes.

Sessions are coordinated by adjuncts instructors and older students which have been previously trained. Usually one group of two students is assigned per bilayer setup. To allow the session to run smoothly and maximize the interactions with the students, one instructor for two groups of students is an optimal configuration. Lab sessions at our institution are run with ten bilayer setups working simultaneously.

\subsection{Equipment}

\subsubsection{Amplifier}

\begin{figure}
\centering
\includegraphics[width=0.7\linewidth]{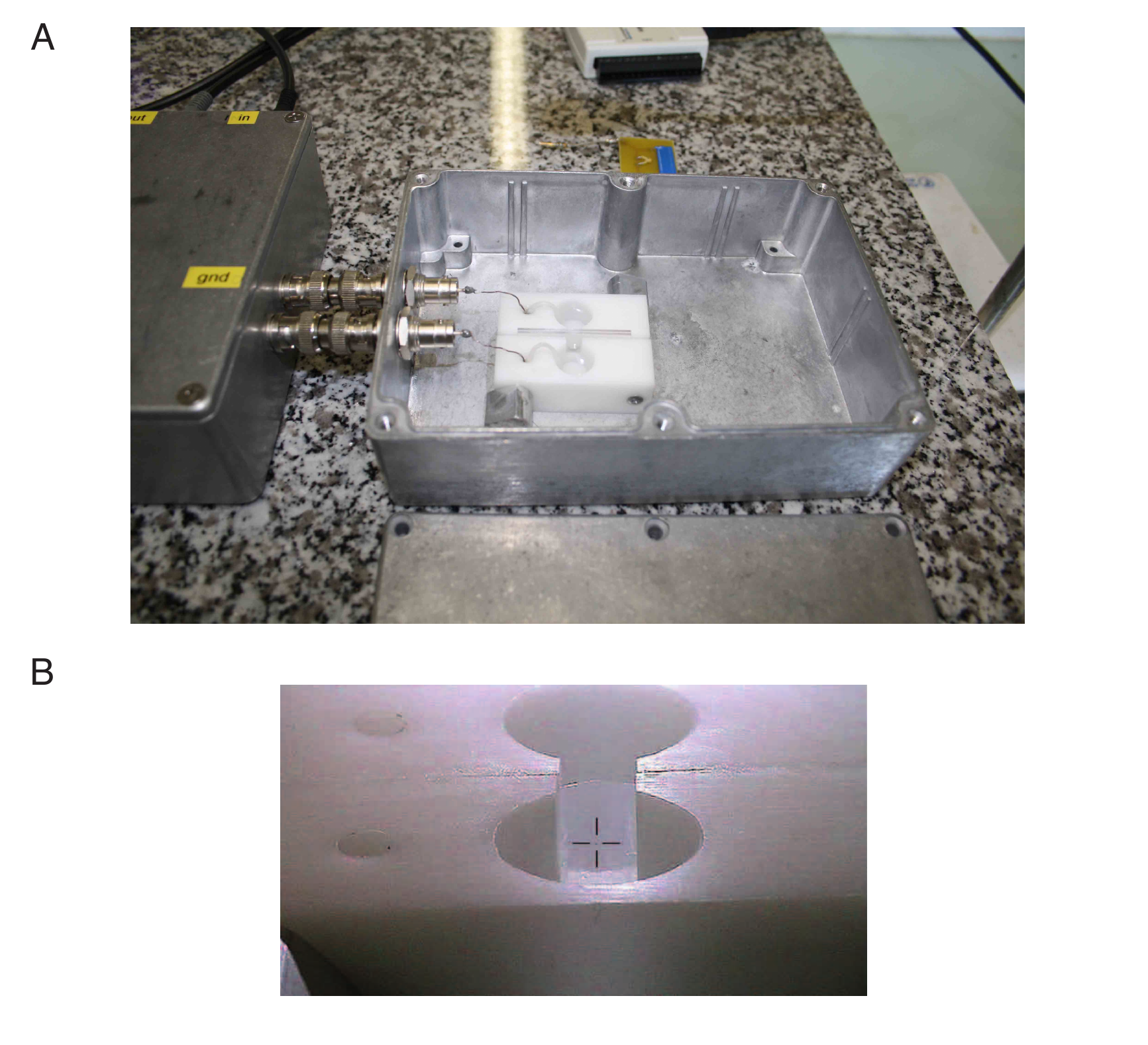}
\caption{Bilayer chamber. (A) The bilayer chamber is sitting in an metallic box that, when closed, acts as a Faraday cage minimizing electric noise. The chamber is composed of two compartments milled in two blocks of polyoxymethylene which are pressed together by a pair of screws. Each compartment is connected to the amplifier though an Ag-AgCl electrode and a salt bridge. (B) the Teflon film is sealed with silicon grease between the two blocks. Before insertion of the film in the chamber, a circular 200 $\mu m$ aperture is first etched in the film (visible at the centre of the cross superimposed on the image).}
\label{BLMchamber-fig}
\end{figure}

We have developed a low cost open-source bilayer amplifier (fig. \ref{OpenPicoAmp-fig}). The amplifier is designed using the lithographic printed circuit board fabrication process and off-the-shelf electronic components, like the TL071 and TL061 operational amplifiers (Texas Instruments, USA). The amplifier head stage is a current to voltage converter equipped with a 1 $G \Omega$ feedback resistor. The two-stage amplifier has two built-in gains of 1 $mV/pA$ and 100 $mV/pA$, low-passed from 3400 to 16 $Hz$ respectively. The amplifier at its highest gain has a sensitivity in the $pA$ range which makes it suitable for the recording of the unitary current produced by the opening of a single channel molecule. The design choices for this amplifier were driven by the following conditions : it should be easy to build and inexpensive. Therefore we have chosen to use commonly found electronic components and kept their number to a minimum. As a consequence this amplifier may not be suitable for research purpose due to bandwidth limitations, but is well adapted to perform the experiment proposed here. The electronics design can be found in the supporting information section of this paper. These design files are licensed under a Creative Commons Attribution Share-Alike license, which allows for both personal and commercial derivative works, as long as this paper is credited and the derivative designs are released under the same license.

\subsubsection{Bilayer chamber and electrical recording}

\begin{figure}
\centering
\includegraphics[width=0.7\linewidth]{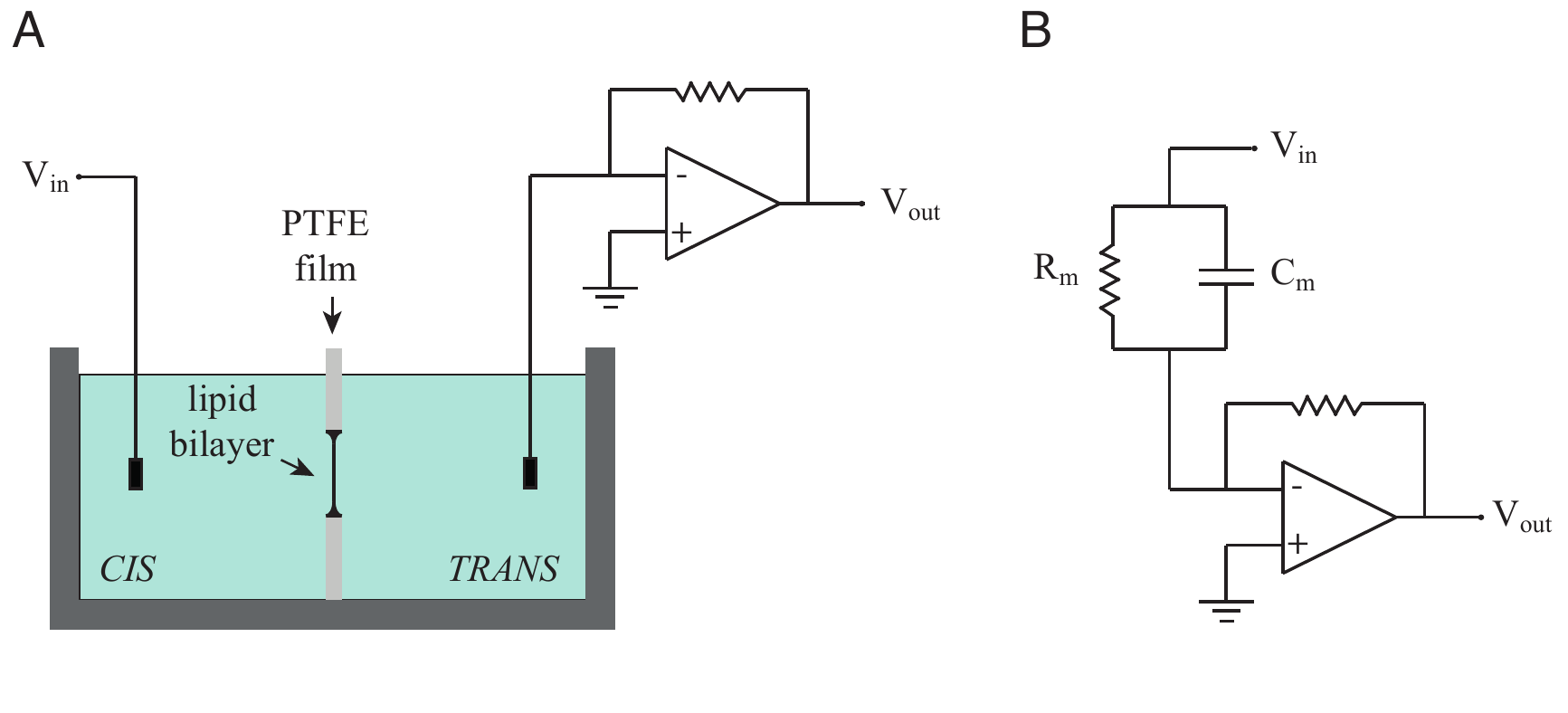}
\caption{The planar lipid bilayer technique. (A) Experimental setup for the electrical recording. The two compartments of the experimental chamber are filled with a 1 $M$ NaCl solution. A small volume of the lipid solution is injected over the aperture in the teflon film and the bilayer is formed by spontaneous thinning. The voltage signal is applied on the $cis$ side while the $trans$ side is held at 0 $V$  by the current to voltage converter, acting as a virtual ground.  (B) A lipid  bilayer is electrically equivalent to leaky capacitor. With the proposed design for the experimental chamber, typical values for the resistance ($R_m$) of a channel-free lipid  bilayer are in the 100 $G \Omega$ range with a capacitance ($C_m$) in the range of hundreds of $pF$. }
\label{circuits-fig}
\end{figure}

\begin{figure}
\centering
\includegraphics[width=0.7\linewidth]{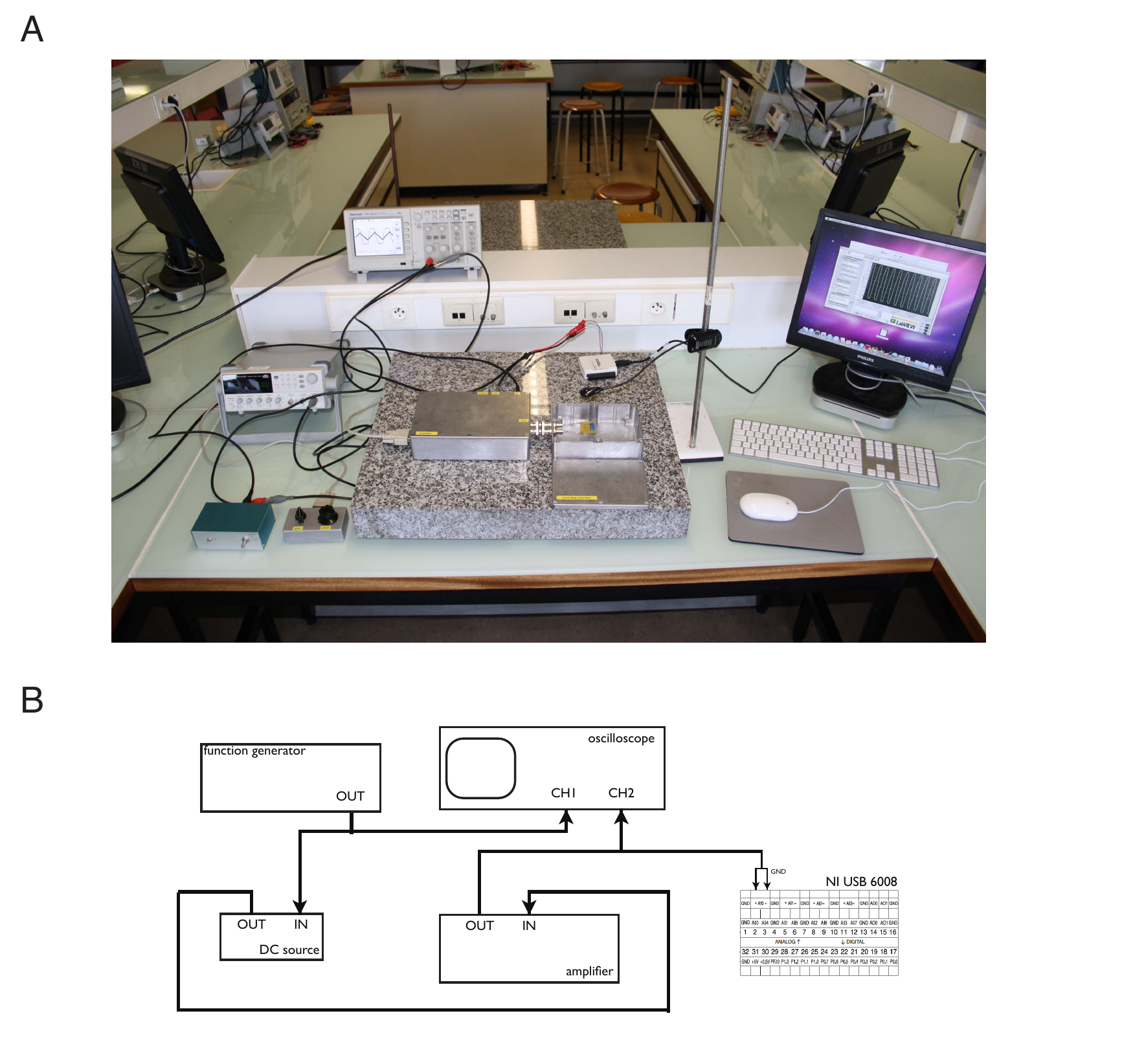}
\caption{The complete planar bilayer setup as implemented at our institution. (A) The different parts of the setup can be seen : the amplifier and its remote control, a waveform generator, a switchable DC source and an oscilloscope. In addition, we use a NI USB6008 for data acquisition with the lab computers. The recording chamber, connected to the main box of the amplifier, is placed on a heavy granite slab resting on four tennis balls. A model circuit composed of a 1 $G\Omega$ resistance with a 220 $pF$ capacitor in parallel is used for calibration. (B) The corresponding block diagram of the setup. }
\label{BLMsetup-fig}
\end{figure}

The bilayer chamber is composed of two compartments milled in two blocks of polyoxymethylene (Delrin$^\circledR$) allowing the insertion of a commercially available polytetrafluoroethylene (PTFE or Teflon$^\circledR$) film (Norton T-100 Premium Grade skived PTFE film, Saint-Gobain, Germany). The films used for this experiment have a thickness of 50 $\mu m$. Before insertion of the film in the chamber, a circular 200 $\mu m$ aperture is first etched in the PTFE film using a modified syringe needle (for more details, see supplemental information). The experimental chamber consists of the Teflon film sealed with silicon grease between the two Delrin blocks which are pressed together by a pair of screws (fig. \ref{BLMchamber-fig}). This design choice allows an easy replacement of the Teflon film which is sometimes teared during the manipulations by the students. The two compartments of the experimental chamber are called $cis$ and $trans$. The $trans$ side is always defined as zero voltage, corresponding to the usual convention in electrophysiology where the reference potential is on the extracellular side (fig. \ref{circuits-fig}.A).  A lipid bilayer, formed in the aperture of the Teflon film is electrically equivalent to a leaky capacitor (fig. \ref{circuits-fig}.B).The experimental chamber and the amplifier circuit are enclosed in an aluminium box that acts as a Faraday cage minimizing electric noise. Mechanical vibrations are damped by placing the experimental chamber on a heavy granite slab resting on four tennis balls. Each compartment is connected to the amplifier though an Ag-AgCl electrode and a salt bridge (1 $M$ KCl, 2\% Agar). The complete setup consists of the amplifier, a waveform generator, a switchable DC source and an oscilloscope. Alternatively, low cost USB interfaces, like the NI USB6008 or the NI MyDAQ (National Instruments, USA), can be used to provide a solution for data acquisition and signal generation using the lab computers (fig. \ref{BLMsetup-fig}).

\subsubsection{Chemicals and solutions}

Commercially available phospholipids can be used to form planar lipid bilayers. We  selected  3-sn-phosphatidylcholine (Sigma-Aldrich, ref. 61758 FLUKA) as soy lipid extracts are the less expensive. The lipids were dissolved in n-octane (Sigma-Aldrich, ref. O2001 SIGMA) to a final concentration of 20 $mg/ml$. This lipid solution has to be prepared shortly before performing the experiments. The salt or electrolyte solutions used in most planar lipid bilayer experiments are often KCl or NaCl. In our case the recordings were made in ionic symmetrical conditions with 1 $M$ NaCl in both compartments or in asymmetrical ionic condition with 1 $M$ NaCl in the $cis$ compartment and 0.1 $M$ NaCl in the $trans$ compartment.

Gramicidin A is a small peptide that forms ionic channels in lipid bilayers \cite{Mueller1967,Hladky1972,Finkelstein1981}. It is active against gram-positive bacteria and used for clinical application as an antibiotic. When two gramicidin A molecules link transiently, they form an open ionic channel which is selective for monovalent cations allowing the passage of millions of ions per second through the lipid bilayer. The corresponding unitary current is therefore in the $pA$ range and can be detected with the proposed amplifier. The duration of this elementary step of current is determined by the life time of the gramicidin dimer extending through the membrane. For the sake of simplicity, we are going to refer to these events as channel openings. These openings typically last several hundreds of $ms$ and are thus also easy to resolve temporally.  Gramicidin A is available commercially and inexpensive (Sigma-Aldrich, ref. 50845 SIGMA). A 1 $nM$ stock solution is prepared by successive dilution with ethanol. The stability of gramicidin A in this solution appears very high (more than 4 years).

\subsection{Experiments}
The laboratory session is divided in three successive steps corresponding to a progressive approach of the different aspects of the topic. The corresponding student hand-out  can be found in the supporting information section of this paper.

\subsubsection{Step 1} The cell membrane is considered to be electrically equivalent to leaky capacitor. Guiding question : How does such a $RC$ circuit reacts when submitted to a variable or continuous voltage signal ? Learning goals : Students will understand that (1) the total current is essentially produced by the capacitive current when the circuit is submitted to a high frequency voltage signal and (2) that the total current is purely resistive when the circuit is submitted to a constant voltage. Therefore these two conditions can be used to actually measure the capacitance and resistance of the circuit elements. Activity : Students have to compute  and measure the capacitive and resistive currents for a circuit composed of a  resistance with a  capacitor in parallel when it is submitted (1) to a high frequency triangular signal and (2) to a constant voltage.  

\subsubsection{Step 2}  Guiding question : How does a lipid bilayer reacts when submitted to similar variable or continuous voltages ? Learning goals : Students will understand that a planar lipid bilayer is electrically equivalent to a $RC$ circuit. Activity : Students have to obtain a lipid bilayer and measure the current first when it is submitted (1) to a high frequency triangular signal and (2) to a constant voltage. This allow them to measure the bilayer capacitance and resistance.

\subsubsection{Step 3}
Guiding question : What is an ionic channel ? Learning goals : Students will observe that certain molecules can form pores in the lipid bilayer allowing transiently the selective passage of ions and therefore act as electrical switches at the molecular level. Activity : Students have to obtain a lipid bilayer in presence of an gramicidin A and measure the unitary currents when the bilayer is submitted to a range of constant voltages.

\subsection{Survey methodology}
At the beginning of the laboratory session, students were given ten minutes to fill out a short answer survey. An additional survey was then administered at the end of session. The pre- and post-surveys consisted of True/False questions about membrane biophysics. The answers were scored as ‘0 = incorrect’and ‘1 = correct’). A ‘‘knowledge score’’ was determined as the sum of correct answers . A paired samples t-test was conducted to compare knowledge scores before and after the demonstration.

\section{Results}

\subsection{Experiments}

In a typical four hours session, all students were able to obtain the lipid bilayer formation and study its electrical properties. In addition, we found that routinely a majority of attendees could actually succeed to observe the single channel currents.  We consider this as a good performance, as it is a rather delicate experiment and for many students this was their first experience in such a laboratory environment.

\subsubsection{Calibration}

Students have to compute the capacitive and resistive currents for a circuit composed of a 1 $G\Omega$ resistance with a 220 $pF$ capacitor in parallel when it is submitted (1) to a 500 $Hz$ 10 $mV$ peak to peak triangular signal and (2) to a constant 120 $mV$ voltage. When the circuit is submitted to the triangular signal, which has a slope of $\pm$ 10 $V/s$, the capacitive current is a square signal with an amplitude of 2200 $pA$. The resistive current is a triangular signal of 10 $pA$ peak to peak amplitude, which is therefore negligible. The total current measured by the amplifier with the 1 $mV/pA$  gain appears therefore a square signal of 2.2 $V$ of amplitude on the oscilloscope which is essentially the steady state capacitive current. When submitted to the constant 120 $mV$ voltage, the capacitive current is zero and there is only a constant resistive current of 120 $pA$ appearing as a 120 $mV$ on the oscilloscope.

Subsequently, they connect the corresponding model $RC$ circuit to the amplifier and do the actual measurements using the oscilloscope. This step also provides a calibration of the amplifier allowing to evaluate the bilayer capacitance during its formation by measuring the capacitive current when the bilayer is submitted to the triangular signal (each $pF$ of capacitance will contribute 10 $mV$ to the amplitude of the signal read on the oscilloscope with amplifier set at the 1 $mV/pA$ gain).

\subsubsection{Bilayer formation}
\begin{figure}
\centering
\includegraphics[width=0.7\linewidth]{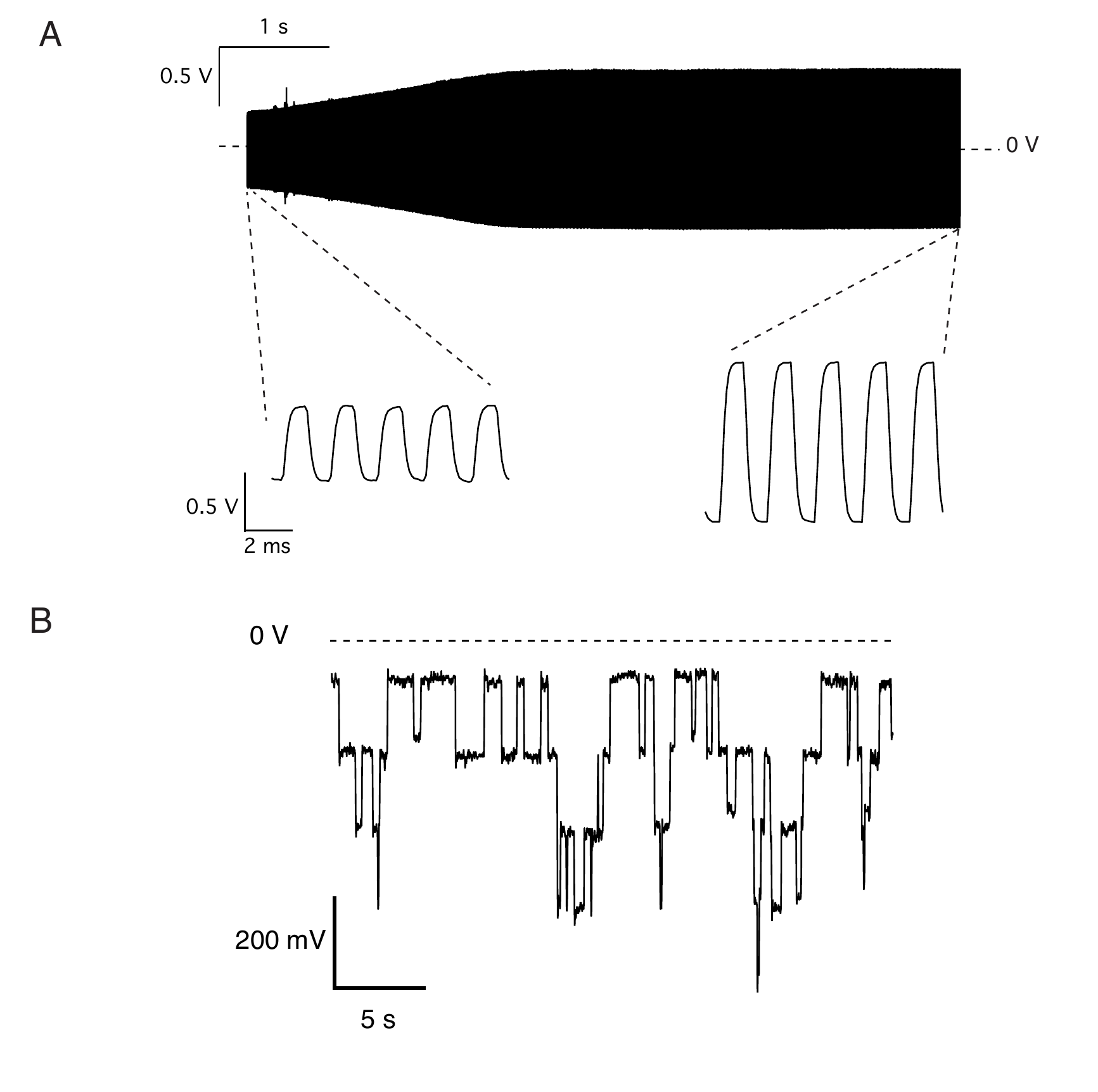}
\caption{Experimental recordings obtained with the OpenPicoAmp. (A). The lipid bilayer formation is monitored by measuring the  increase in the capacitive current during the application of a triangular voltage signal (500 $Hz$, 10 $mV$ peak to peak). In this particular example, the lipid bilayer forms and stabilizes in less than 10 $s$ with a final capacitance around 70 $pF$. (B)  After addition of gramicidin A, the students are able to record unitary currents when the bilayer is submitted to constant voltage. In the presented recording, obtained at -120 $mV$, a maximum of four gramicidin A pores are open at the same time. All currents traces appears as voltage deflections as the amplifier acts as a current to voltage converter, with a gain of 1 $mV/pA$  (panel A) or 100 $mV/pA$  (panel B). All the presented data has been acquired by students using a National Instruments USB6008 A/D converter using a sampling frequency of 10 $kHz$ and a low-pass filter cutoff at 3.4 $kHz$ (panel A) or a 100 $Hz$ sampling frequency with a low-pass filter cutoff at 16 $Hz$ (panel B).}
\label{BLMrec-fig}
\end{figure}

The two compartments of the experimental chamber are first filled with electrolyte solution (1 $M$ NaCl). Subsequently 2 $\mu l$ of a 20 $mg/ml$ solution of 3-sn-phosphatidylcholine dissolved in n-octane are injected over the aperture using a 10 $\mu l$ pipette (see movie in the supplementary information). The bilayer is formed by spontaneous thinning. If the lipid bilayer does not form within a few minutes after adding the lipids, the electrolyte solution is agitated to speed up the thinning process. The agitation can be achieved by producing small air bubbles near the aperture with a 10 $\mu l$ pipette. The bilayer formation is monitored by measuring the capacitive current  during the application of a triangular voltage signal (500 $Hz$, 10 $mV$ peak to peak). As a the planar lipid bilayer is forming in the central part of the aperture, the capacitive current will increase and then stabilize (fig. \ref{BLMrec-fig}.A). A stable lipid bilayer formed inside the 200 $\mu m$ aperture in the teflon film will typically produce a capacitance in the range of hundreds of $pF$. The capacitance of the bilayer is the first criteria for proceeding with an experiment. Low capacitance reflects a bilayer that has not completely formed or that is overly thick. In that case further agitation is needed to speed up the thinning process. The final bilayer stability measurement is the amount of current flowing across a channel-free bilayer in response to test potentials in the voltage range that is to be used in the experiment. If test potentials generate substantial currents across the membrane in the absence of ionic channels, the bilayer is deemed unstable and should be broken and reformed before beginning an experiment. In our case, the total current at a constant 120 $mV$ voltage is measured and should be below 2 $pA$.

Knowing the specific capacitance of a solvent-containing lipid bilayer (0.5 $\mu F/cm^2$) and the intensity of the capacitive current, the students are also able to estimate the bilayer diameter which should be slightly lower than the aperture diameter due to the presence of the annulus at the periphery of the bilayer \cite{White1972}. In addition, by measuring the membrane current while applying a constant holding potential and knowing that the typical thickness of a planar lipid bilayer is around 10 $nm$, they can provide an estimate of the  bilayer resistivity and demonstrate that lipids act as insulators as the typical value is around $10^3$ $G\Omega .cm$. This experiment allows the students to understand that the electrical response of a passive cell membrane is well described by an equivalent $RC$ circuit. There is no difference in the electrical response between the model circuit and the lipid bilayer.

\subsubsection{Unitary currents}
\begin{figure}
\centering
\includegraphics[width=0.7\linewidth]{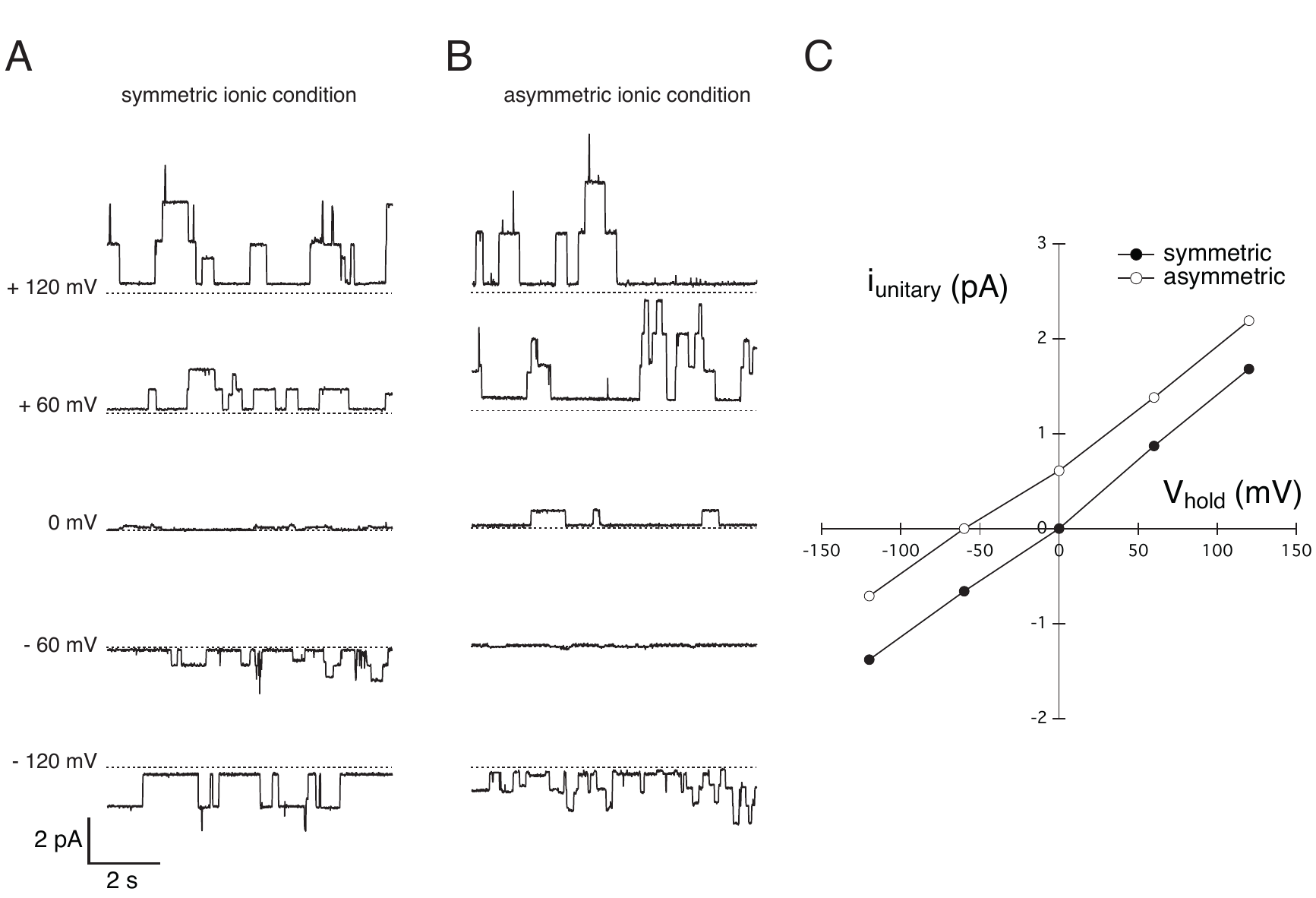}
\caption{Effect of ionic gradients on gramicidin A unitary currents, recorded with the OpenPicoAmp. (A) Gramicidin A unitary currents in symmetric ionic conditions (1M $NaCl$ $cis$ / 1 M $NaCl$ $trans$) at different holding potentials. The dotted line on each individual recording corresponds to 0 $pA$. (B) Gramicidin A unitary currents in asymmetric ionic conditions (1M $NaCl$ $cis$ / 0.1 M $NaCl$ $trans$) at different holding potentials. (C) Gramicidin channels are ohmic conductors. The unitary conductance observed in these recordings is around 13 $pS$. The reversal potential of the unitary currents is shifted from 0 $mV$, in symmetric ionic condition, to approximatively -60 $mV$, in asymmetric ionic condition.  All data was acquired using a NI USB 6008 A/D converter with a 100 Hz sampling frequency with the amplifier gain at 100 $mV/pA$ (low-pass filter cutoff at 16 $Hz$).}
\label{sym_asym-fig}
\end{figure}

After breaking the previously obtained channel-free bilayer, students are asked to reformed it in the presence of 1 $pM$ of the ionophore gramicidin A in each compartment. This low concentration is chosen to obtain less than ten simultaneous individual channel openings, allowing the resolution of the elementary current produced by the transient dimerization of two gramicidin A molecules. After addition of gramicidin A to the aqueous phase it may require up to 15 minutes for the molecules to diffuse to the bilayer and insert. The students are able to record unitary currents when the bilayer is submitted to constant voltage (fig. \ref{BLMrec-fig}.B). This further allows the study of the biophysical properties of the gramicidin pores in symmetric ionic condition (1M $NaCl$ $cis$ / 0.1 M $NaCl$ $trans$) by measuring the unitary currents, $i$, at different values of the holding potential (fig. \ref{sym_asym-fig}.A). The current mediated by each channel is proportional to the applied voltage, demonstrating the ohmic behavior of the gramicidin channels :

  \begin{equation}
	i=\gamma(V_{hold}-V_{Nernst})
 \end{equation}

\noindent where $\gamma$ is the conductance of a single gramicidin channel, $V_{hold}$ the applied transmembrane potential and $V_{Nernst}$ the reversal potential for $Na^+$ ions, which is given by the Nernst equation:

 \begin{equation}
	 V_{Nernst}=\frac{RT}{F}ln\frac{[Na^+]_{trans}}{[Na^+]_{cis}}
	 \label{nernst1-eqn}
 \end{equation}
 
\noindent where $R$ is  universal gas constant, $T$ the absolute temperature, $F$ the Faraday constant, $[Na^+]_{trans}$ and $[Na^+]_{cis}$ the $Na^+$ concentrations in the $trans$ and $cis$ compartments respectively. Therefore when these measures are  repeated in asymmetric ionic conditions (1M $NaCl$ $cis$ / 0.1 M $NaCl$ $trans$), the reversal potential of the unitary current is shifted by approximatively 60 $mV$ compared to the symmetric ionic condition (fig. \ref{sym_asym-fig}.B and \ref{sym_asym-fig}.C). 

\subsection{Survey results}
We assessed the improvement in the student knowledge of basic biophysics by comparing the knowledge scores (total points) that were collected both before and after the demonstration. There was a significant difference (p = 0.0015, n=121) in the pre-test (M = 7.12, SD = 1.79) and the post-test knowledge (M = 7.85, SD = 1.77), suggesting that the proposed material improved their knowledge of core concepts of membrane biophysics.


\section{Discussion}

Understanding the human brain is one of the greatest challenges facing contemporary science. Through the progress of neuroscience, we can gain profound insights into our inner workings, develop new treatments for brain diseases and build revolutionary new computing technologies.  Besides the Obama administration’s Brain Initiative and the European Union’s Human Brain Project, there are numerous private and public research efforts in Europe and abroad, focusing on the human brain.
This will undoubtedly provide a surge of activity in brain research as scientists try to build the tools and knowledge they need to accomplish their goals. Therefore it is of particular importance to promote neuroscience education to fuel this initiatives by providing inspiration to future scientists.

In this paper, we provide a basic experimental protocol allowing university undergraduate students to build artificial cell membranes and examine ionic channels properties at the single-molecule level. In this framework, we have developed an open-source lipid bilayer amplifier, the OpenPicoAmp, which is appropriate for use in introductory courses in biophysics or neurosciences. Such a low cost, open source and well documented solution to build a complete planar lipid bilayer setup was not available so far, although similar projects exist for extracellular recordings of neural electrical activity \cite{Marzullo2012,Land2001}. This type of open-source hardware projects will without doubt encourage collaborative tinkering, promote science education and may even prove to be useful as research tools \cite{Pearce2012, nmeth2013}.

The proposed experimental protocol provides a simple example and can be further refined if the students attend more laboratory sessions. For example, the single channel recordings with gramicidin can be repeated in the presence of pharmacological agents affecting the gating of the channels, e.g. 50 mM fluoxetine \cite{Kapoor2008}. In addition, other ionic channels could be studied but, in this case, the electronics of the amplifier may need to be adapted to allow the recordings of faster events. More advanced subjects can also be introduced, like the determination of single channel current and mean channel opening times using autocorrelation function analysis of current noise  \cite{Kolb1977}. The design of the bilayer chamber could also be adapted to allow the simultaneous optical monitoring of the bilayer formation. It is possible to view the phospholipid film using a lens with 10 x magnification and a light source. The thin film interference of the reflected light at the water/membrane and membrane/water interfaces produces colorful  fringes when the membrane thickness is on the order of the wavelength of light in the membrane. Once the film reaches nanometer-scale thickness, the membrane turns black because light reflected off the two interfaces interferes destructively for all wavelengths; hence the name black lipid membrane (BLM) \cite{TIEN1966}. Finally, at this stage, we do not provide any open source software for data acquisition and visualization. But such software exists \cite{Dempster2001} and we are currently developing a solution that will be tailored to the experimental protocol proposed here.

The reactions of the students to the lab sessions are positive and their learning outcomes are clearly improved by this hands-on approach. We hope that the proposed experiment will be used as a teaching tool in other institutions, allowing fruitful exposure of undergraduate students to biophysics or neurosciences.

\section{Supporting information}

\subsection{Electronics}
\begin{itemize}
\item {\bf File S1} : CadSoft Eagle .brd file giving the board layout for printing amplifier circuit. The CadSoft Eagle software Light Edition (freeware) can be downloaded from \url{http://www.cadsoftusa.com/}.
\item {\bf File S2} : CadSoft Eagle .sch file including  the circuit diagram for the amplifier circuit and the linked .brd file.
\item {\bf File S3} : A PDF document detailing the practical implementation of the amplifier, the parts and costs of components with associated part numbers from an electronics supplier. A ready to print layout for the printed-circuit board of the OpenPicoAmp is also given (scale 1:1).

\end{itemize}

\subsection{Bilayer chamber}
\begin{itemize}
\item {\bf File S4} : An  PDF document detailing the design of our bilayer chamber (all dimensions given in $mm$).
\item {\bf File S5} : A  PDF document describing the preparation of the Teflon film, the assembly of the bilayer chamber and the making of the electrodes.
\end{itemize}

\subsection{Video and hand-outs}

\begin{itemize}
\item {\bf File S6} : A  short video .avi showing the proposed experiment (download link: \url{http://homepages.ulb.ac.be/~dgall/S6_BLM.avi})

\item {\bf File S7} : The student hand-out for the experiments (PDF).

\item {\bf File S8} : The student survey questions (PDF).

\end{itemize}


\section*{Acknowledgments}
We thank Prof. Fabrice Homblé for careful reading of the manuscript, Prof. Francis Grenez for supporting the project, Tiziano D'Angelo and Emmanuel Hortmanns for their technical help. This project has received support from the Fonds d'Encouragement à l'Enseignement  of the Université Libre de Bruxelles.

\bibliographystyle{hplain}
\bibliography{DGnew}

\


\end{document}